# Multimodal surface coils for low-field MR imaging

Yunkun Zhao[1], Aditya A Bhosale[1], Xiaoliang Zhang[1,2*]

[1]Department of Biomedical Engineering, [2]Department of Electrical Engineering, State University of New York at Buffalo, Buffalo, NY, United States

*Corresponding author:

Xiaoliang Zhang, Ph.D.
Bonner Hall 215E
Department of Biomedical Engineering
State University of New York at Buffalo
Buffalo, NY, 14226
U.S.A.

Email: xzhang89@buffalo.edu

**Abstract**

Leveraging the potential of low-field Magnetic Resonance Imaging (MRI), our study introduces the multimodal surface RF coil, a design tailored to overcome the limitations of conventional coils in this context. The inherent challenges of low-field MRI, notably suboptimal signal-to-noise ratio (SNR) and the need for specialized RF coils, are effectively addressed by our novel design. The multimodal surface coil is characterized by a unique assembly of resonators, optimized for both B1 efficiency and low-frequency tuning capabilities, essential for low-field applications. This paper provides a thorough investigation of the conceptual framework, design intricacies, and bench test validation of the multimodal surface coil. Through detailed simulations and comparative analyses, we demonstrate its superior performance in terms of B1 field efficiency, outperforming conventional surface coils.

**Introduction**

Magnetic resonance imaging (MRI) is a powerful tool in the medical field, known for its ability to provide detailed images of the internal structures, functions and metabolic processes of the living system without the use of ionizing radiation [1-18]. Traditional high-field MRI systems, typically operating at 1.5 Tesla or higher, are prevalent in clinical settings due to their high signal-to-noise ratio (SNR), which contributes to their ability to produce high-resolution images [19-32]. However, these systems are not without their limitations. High-field MRI systems are associated with high operating costs, substantial power requirements, radio frequency (RF) challenges [33-37] and safety concerns, especially for patients with certain medical implants or conditions that contraindicate exposure to strong magnetic fields [38-49]. In recent years, there has been a growing interest in low-field MRI systems, defined as those operating at magnetic field strengths in the range from 0.25 Tesla to 1.0 Tesla. These systems offer several advantages over their high-field counterparts, including lower operating costs, reduced power consumption, and improved safety profile, making them more accessible and suitable for a wider range of patients [50, 51].

However, a significant challenge associated with low-field MRI is the inherently low signal-to-noise ratio (SNR) [52-55]. SNR is a critical factor in MRI, as it directly affects the quality and resolution of the images obtained [56-58]. One crucial aspect of improving the quality of MRI, especially in low-field systems, is the enhancement of Radio Frequency (RF) coil designs. The RF coil is a key component in MRI systems, responsible for transmitting and receiving the magnetic signals that form the basis of the MRI images. Improving the efficiency and performance of these coils can lead to significant advancements in MRI quality, particularly in terms of SNR and image resolution.

The SNR in MRI is known to be directly related to the B1 efficiency of the RF coils used in the imaging process. Conventional surface coils, which offer the highest B1 efficiency among all RF coil types, are often unable to provide sufficient transmit/receive field strength in low field MR applications, resulting in suboptimal image sensitivity and resolution. Therefore, the development of more efficient RF coils should provide a tangible solution to improve imaging performance at low fields. To address this challenge, in this work we introduce and investigate an innovative solution: the multimodal surface coil. This novel design significantly improves the B1 fields over those provided by conventional surface coils, with the potential to significantly enhance MR SNR. The core concept of the multimodal surface coil is based on a set of stacked resonators, which are electromagnetically coupled to form a multimodal RF resonator. This design not only improves B1 efficiency but also incorporates a low-frequency tuning capability. Such a feature is particularly advantageous in low-field MRI, where tuning at the corresponding low frequencies poses a notable challenge. To validate our design, we have conducted extensive full-wave electromagnetic simulations alongside standard RF bench tests and measurements. The proposed design is further validated by a comparison study with a conventional surface coil.

## Methods

*EM simulation*

Figure 1 displays the simulation model of a multimodal surface coil. This coil is comprised of seven resonators or coil loops constructed with 6.35 mm wide copper tape. Six of these are identical square coils, each with a side length of 10 cm and equipped with a 60pF capacitance tuning capacitor for tuning to 42 MHz. The central coil contains an impedance matching circuit essential for driving the multimodal surface coil. A spacing of 5 mm between the coils is designed to enhance mutual inductive coupling, with the entire stacked assembly reaching a height of 3 cm. Our design, featuring multiple coils, inherently supports four resonant modes within the coupled stack. We have chosen to utilize the lowest resonant mode for imaging applications due to its superior field strength efficiency, a critical factor in optimizing image quality in MRI. A conventional surface coil has been used as a comparative setup. Mirroring the multimodal resonator, it was also defined by a 10 cm length. All designs, including the multimodal surface coil, were built using a 6.35 mm wide copper sheet conductor, tuned to 21.3 MHz, and impedance-matched to 50 ohms. In comparison study, a square air phantom has been placed 1 cm above the coil with 5 cm height and 10 cm length and width as an imaging area for B1 field strength and distribution evaluation. Performance assessments of the multimodal surface coil involved analyzing scattering parameters, and B1 efficiency, using field distribution plots. All electromagnetic field plots were normalized to 1 W of accepted power. Numerical results of the proposed designs were obtained using the electromagnetic simulation software CST Studio Suite (Dassault Systèmes, Paris, France).

*Bench Test Model Assembly*

Figures 2 shows photographs and dimensions of bench test models of multimodal surface coil and conventional surface coils. The bench test models have the same dimensions as the simulation model. The conductors of the multimodal surface coils were built with 6.35 mm wide copper tape and on a 3D printed polylactide structure. The imaging resonant frequency was tuned to 21 MHz and matched to 50 ohms by careful selection of the capacitance value on each individual coil. We used 7 identical fixed tuning capacitors with 39 pF capacitance. The matching circuit was implemented as shown in Figure 2A. One capacitor with 330 pF connected in parallel to the feeding line was employed for impedance matching.

For comparison, a conventional surface coil has also been made. The conventional surface coil has same dimensions as their simulation model and the multimodal surface coil. It was also built using 6.35 mm width thick copper tape on 3D printed polylactide structure and was tuned to 21 MHz and matched to 50 ohms by tuning capacitors and a matching circuit. The B1 field strength and distribution has been visualized by a sniffer positioning system combined with a magnetic and electric field measurement setup shown in Figure 3.

**Results**

*Simulated Resonant Frequency and Field Distribution*

Figure 4 presents the simulated scattering parameters vs frequency for the multimodal surface coil. This graph indicates the emergence of strong coupling between the coils, which manifests as four distinct split resonant peaks. The highest of these peaks occurs at 79 MHz, while the lowest, which is pertinent for imaging purposes, is observed at 21 MHz. It is noteworthy that each individual coil in the originally resonates at 42 MHz, underscoring the effectiveness of our tuning approach using relatively low capacitance value to achieve lower resonant frequency for coil with same size. As comparison, a conventional surface coil with same size requires a 200pF capacitance to reach 21 MHz, leading less accurate and convenience frequency tuning. In Figure 5, we illustrate the B1 field efficiency maps across Y-Z, X-Z, and X-Y planes within the phantom, positioned at the center of the axis. These plots visualize the B1 field efficiency of the multimodal surface coils. The results from these simulations demonstrate that our multimodal surface coil exhibits a stronger B field efficiency, while maintaining a B1 field distribution comparable to that of a conventional surface coil.

*Measured Scattering Parameters and Field Distribution*

Figure 6 shows that the S-parameter vs. frequency plots of the coupled stack-up coil is in good agreement with the simulation results. Four resonant modes with 21.8 MHz, 58.9 MHz, 73.6 MHz, and 85.4 MHz were formed. Figure 7 shows the B1 field efficiency distribution map on X-Y, Y-Z, and X-Z plane measured with 3-D magnetic field mapping system. Multimodal surface coil shows strong B field efficiency and similar field distribution pattern on all three planes and is in accordance with the simulation result, which also indicates that the simulation results are accurate and reliable.

*Field Distribution and Efficiency Evaluation*

Figure 8 provides a comparative analysis of the simulated B1 field efficiency between the multimodal surface coil, and a conventional surface coil. The results clearly demonstrate that, despite maintaining a field distribution similar to that of the conventional surface coil, the multimodal surface coil exhibits a significantly higher B1 field efficiency. This finding is crucial as it underscores the effectiveness of the multimodal design in improving the quality of MRI imaging. In Figure 7B, we present a 1-D plot of the B1 field efficiency, correlating to the vertical and horizontal dashed lines depicted in Figure 7A. This graphical representation provides a detailed insight into the spatial efficiency of the B1 field. Notably, the average B1 field efficiency generated by the multimodal surface coil surpasses that of the conventional surface coil by 46.8% along the X-axis, measured at a position 2 cm above the coil.

Figure 9 presents the comparison of the B field efficiency between the bench test model of the multimodal surface coil and the conventional coil. The measured B-field efficiency distribution and strength are found to be in alignment with our simulation results. This congruence is vital as it validates our design, confirming that the multimodal surface coil possesses a stronger B1 field efficiency within the imaging area compared to the conventional surface coil.

**Discussion**

A key aspect of our study that warrants further exploration is the potential for reducing the overall height of the multimodal surface coil. This can be achieved through more precise fabrication techniques and by minimizing the gap between individual coils. Such refinements in the coil design are not only technologically feasible but also hold significant implications for the coil's performance and application. Reducing the inter-coil spacing, in accordance with the principles of inductive coupling, would likely result in a corresponding decrease in the capacitance value required for each coil. This adjustment could lead to a more compact coil design, enhancing the practicality and usability of the coil in various MRI settings.

Moreover, a reduction in coil spacing is anticipated to positively impact the magnetic efficiency of the system. Smaller gaps between the coils would facilitate stronger mutual inductive coupling, potentially leading to an increase in the B1 field efficiency. This improvement is especially relevant in the context of low-field MRI, where maximizing the efficiency of the B1 field is crucial for achieving high-quality imaging results. The expected enhancement in magnetic efficiency with a more compact coil design could thus represent a significant advancement in low-field MRI technology.

Additionally, these proposed modifications in the coil design may open new direction for further research and development. For instance, exploring the effects of varying the size and spacing of the coils on the resonant frequencies and field distribution could provide deeper insights into optimizing coil designs for specific imaging applications. Such investigations would contribute to a more nuanced understanding of the interplay between coil geometry, electromagnetic properties, and imaging performance, ultimately leading to the development of more efficient and versatile RF coil systems.

Exploring the potential of constructing an array using our multimodal surface coils presents an exciting avenue for advancing MRI technology. The magnetic wall decoupling technique, or induced current elimination (ICE) can be used to reduce electromagnetic coupling between adjacent coils but still preserve the coupling inside the multimodal surface coil, which can be used for building a multimodal surface coil array. Such an array could significantly improve coverage and signal-to-noise ratio across a larger imaging area, while mitigating the challenges of mutual interference between coils. This advancement not only promises to augment field strength and uniformity but also offers potential improvements in parallel imaging, paving the way for faster, more efficient MRI scans.

**Conclusion**

In conclusion, the proposed multimodal surface RF coil has demonstrated significant improvements in B1 efficiency over conventional surface coils, ultimately leading to an improved SNR for low-field MRI. The proposed multimodal surface coil design also helps to achieve low frequency tuning which is technically challenging at low magnetic fields. This multimodal surface coil technique can be possibly used to construct multichannel RF coil arrays when appropriate decoupling techniques are employed, or to design high-field RF coils for small animals.

**Acknowledgments**

This work is supported in part by the NIH under a BRP grant U01 EB023829 and by the State University of New York (SUNY) under SUNY Empire Innovation Professorship Award.

**Figures**

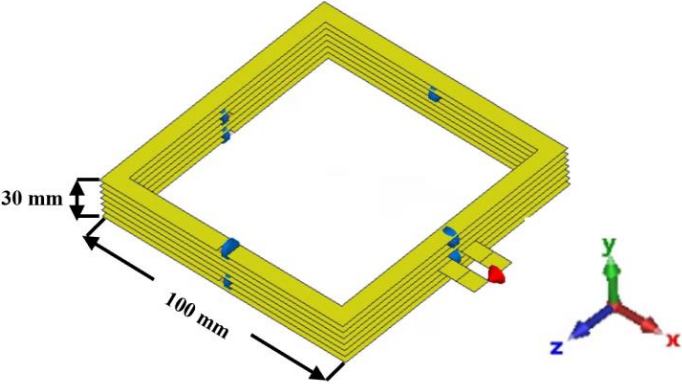

Figure 1
Simulation model and size of multimodal surface coil.

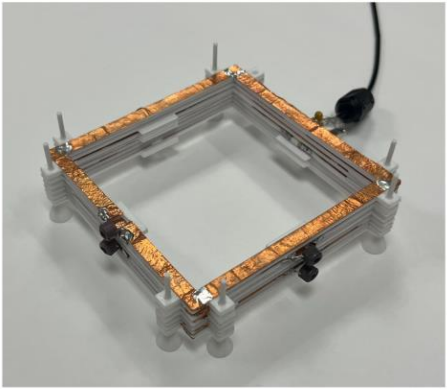 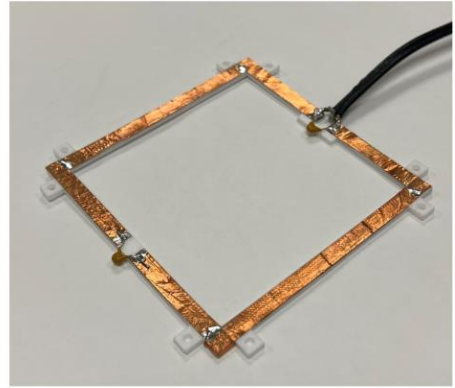

Figure 2
(A) Bench test model of multimodal surface coil. (B) Bench test model of conventional surface coil.

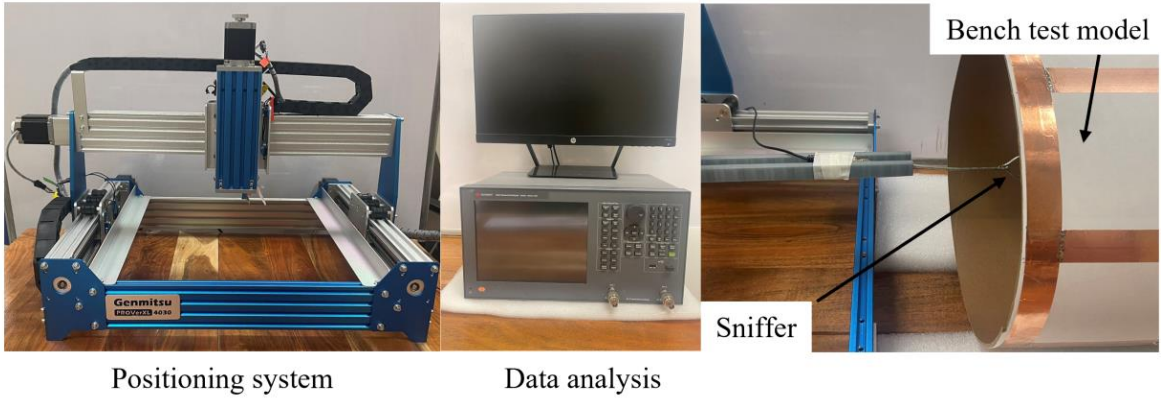

Figure 3
Experimental setup of the sniffer-positioning system combined magnetic field measurement.

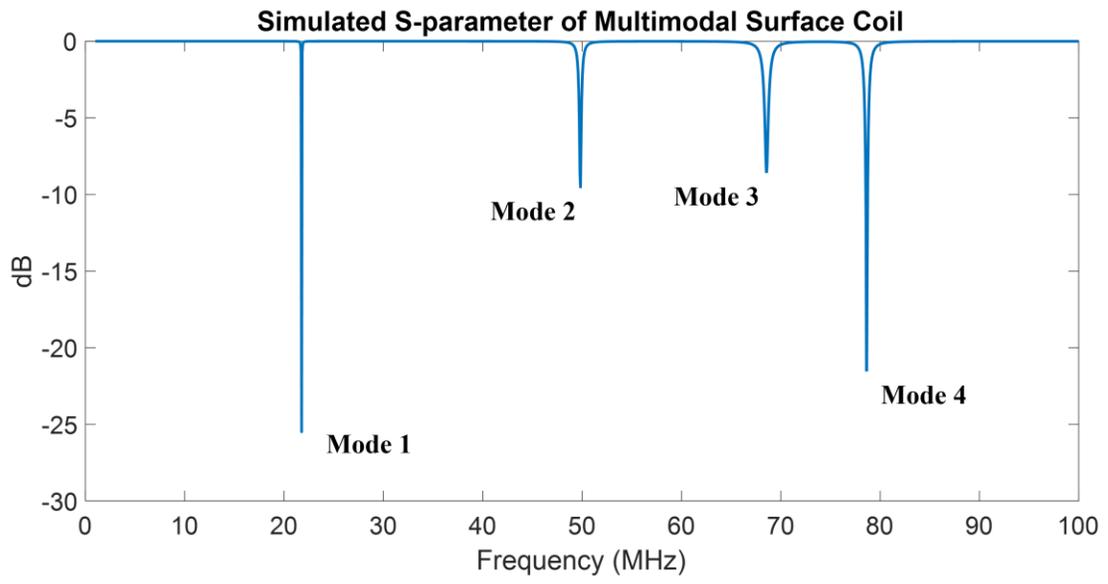

Figure 4
Simulated scattering parameters vs. frequency of the multimodal surface coil.

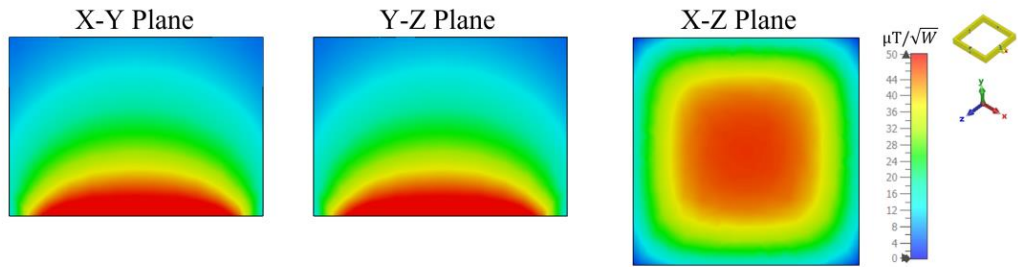

Figure 5
Simulated Y-Z, X-Y, and X-Z plane B field efficiency maps inside phantom generated by multimodal surface coils

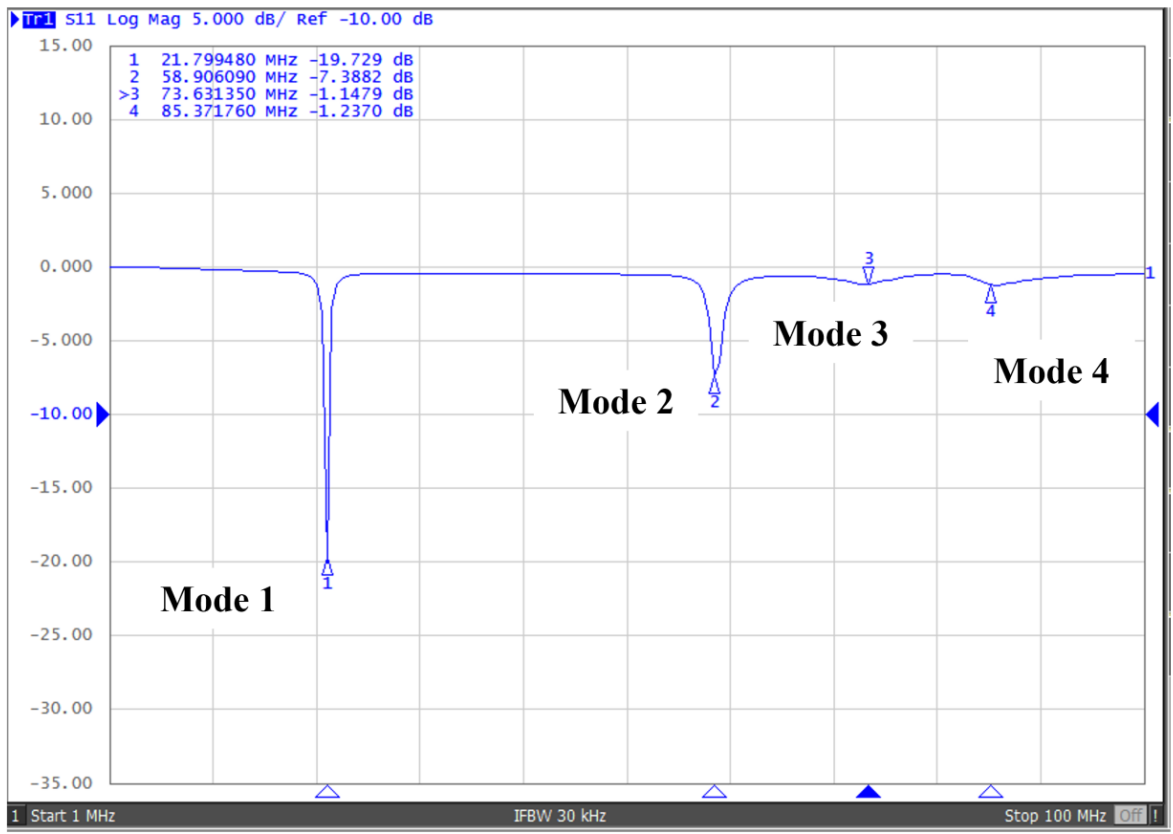

Figure 6
Scattering parameters vs. frequency of the bench test model of multimodal surface coils.

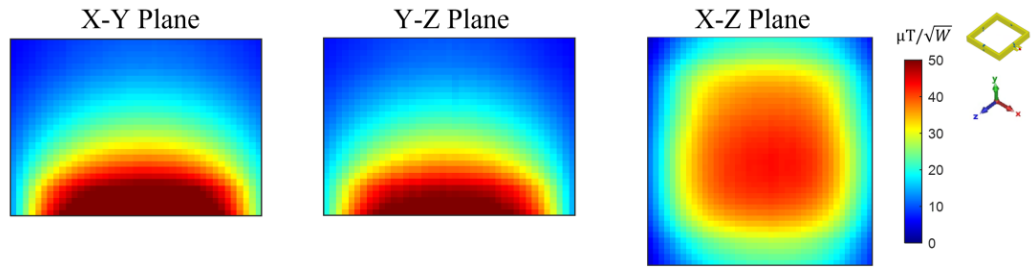

Figure 7
Measured B field efficiency maps on the X-Y, Y-Z, and X-Z plane of multimodal surface coil.

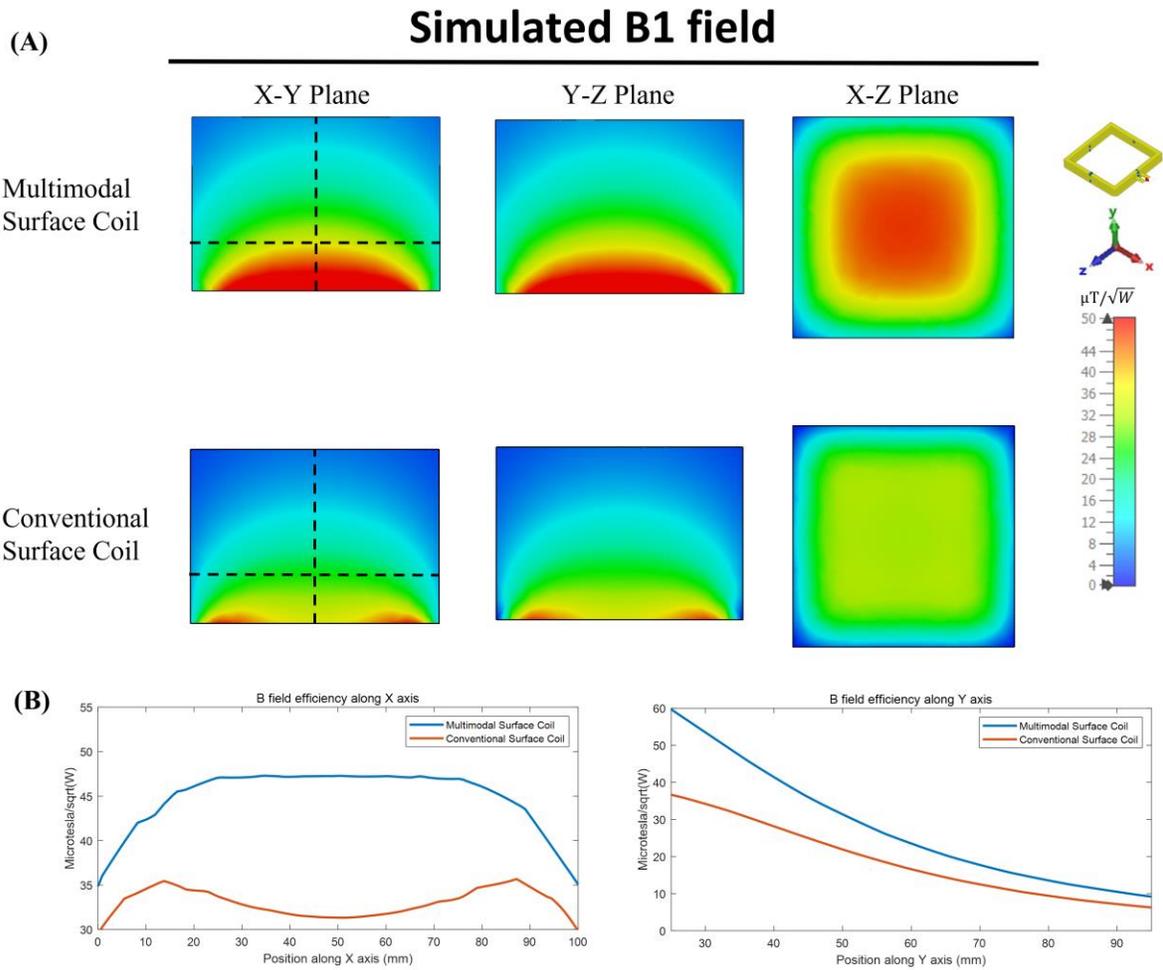

Figure 8
(A) Simulated Y-Z, X-Y, and X-Z plane B field efficiency maps inside phantom generated by multimodal surface coils, and conventional surface coil. Y-Z and X-Y planes are at center of the coil and X-Z plane is at 2 cm above the coil. (B) 1-D plot of B1 field efficiency along the vertical and horizontal dashed line shown in (A).

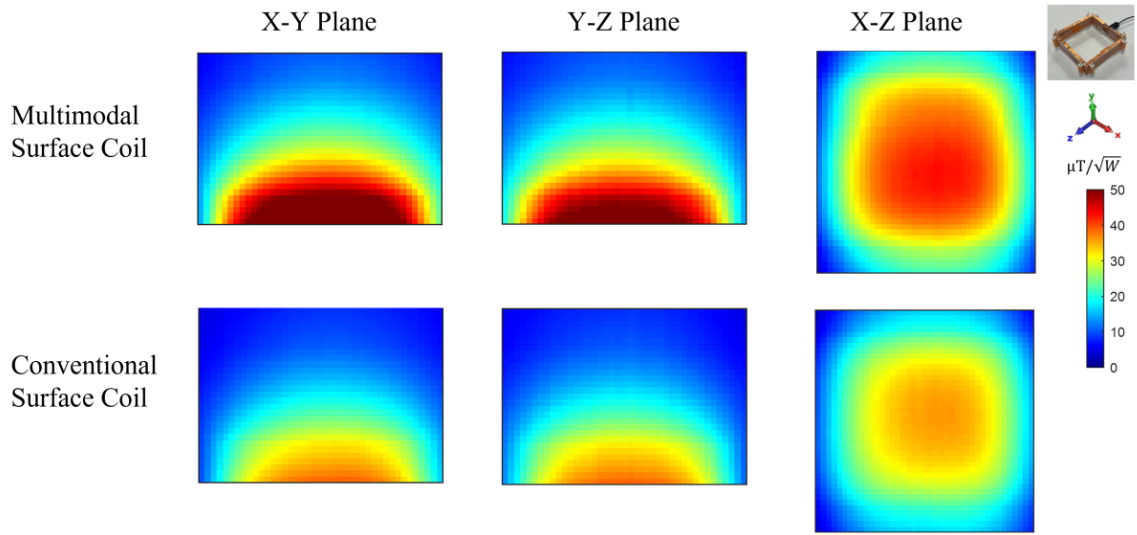

Figure 9
Comparison between measured B field efficiency maps on the X-Y, Y-Z, and X-Z plane of multimodal surface coils and conventional surface coil.